\documentclass[prl,floats,twocolumn,aps,superscriptaddress,showpacs]{revtex4}

\usepackage{amsmath}
\usepackage{graphicx}
\usepackage{color}
\newcommand{\forget}[1]{}

\begin{document}

\title{Capillary  filling with randomly coated walls}

\date{\today}

\author{Fabiana Diotallevi}
\affiliation{Istituto per le Applicazioni del Calcolo CNR 
   V. Policlinico 137, 00161 Roma, Italy}

\author{Andrea Puglisi}
\affiliation{CNISM and Dipartimento di Fisica, Universit\`a Sapienza, p.le A. Moro 2, 00185 Roma, Italy}

\author{Antonio Lamura}
\affiliation{Istituto per le Applicazioni del Calcolo CNR, Via Amendola 122/D, 70126  Bari, Italy}

\author{Sauro Succi}
\affiliation{Istituto per le Applicazioni del Calcolo CNR 
   V. Policlinico 137, 00161 Roma, Italy}

\begin{abstract} 
  The motion of an air-fluid interface through an irregularly coated
  capillary is studied by analysing the Lucas-Washburn equation with a
  random capillary force. The pinning probability goes from zero to a
  maximum value, as the interface slows down. Under a critical
  velocity, the distribution of waiting times $\tau$ displays a
  power-law tail $\sim\tau^{-2}$, which corresponds to a strongly
  intermittent dynamics, also observed in experiments.  We elaborate a
  procedure to predict quantities of experimental interest, such as
  the average interface trajectory and the distribution of pinning
  lengths.

\end{abstract}

\maketitle

The ever-growing technological capability of shaping-up new micro-devices has
revived a keen interest of the scientific community towards the problem
of a liquid-vapor contact line moving on solid
surfaces~\cite{degennes85}. This is a widely studied phenomenon, which
presents subtle effects at different length-scales, challenging
hydrodynamics, thermodynamics, and non-equilibrium statistical
mechanics. At the same time, this issue provides a case-study for a
whole range of industrial applications, where few properties of the
system, e.g. surface smoothness or chemical coating patterns, can be
tuned in order to achieve the desired imbibition efficiency. Other authors
in the past have studied the evolution of the contact line on a
heterogeneous surface, focusing on the deformation of the line along
transversal directions \cite{cox83,joanny84,yeomans_drops}.  Our aim here is to
provide both qualitative and quantitative results for the case of a
narrow capillary with non homogeneous walls \cite{yeomans,noi_pre}.
To this purpose, we focus on the
dynamics of the interface midpoint only, all other details being
projected out through the introduction of a position-dependent
capillary force.  The specific source of irregularity is not crucial (being it
wall roughness or random chemical coating) as long as it can be
described in terms of a fluctuating capillary force experienced by the
fluid-vapor interface. An important difference with previous works is
the inclusion of all inertial and viscous effects, which make the
problem highly non-linear, thus leading to non trivial scenarios even in
the simple case of finite memory randomness.

The Lucas-Washburn equation~\cite{washburn,Lucas} is a credited
model to describe the dynamics of a fluid penetrating an empty
capillary~\cite{tas,washburn_rec}, whose interface midpoint $z(t)$
obeys the dynamic equation:
\begin{equation} \label{lw}
z\frac{d^2z}{dt^2}+\left(\frac{dz}{dt} \right)^2=-\eta z \frac{dz}{dt}+f(z),
\end{equation}
where, in 2D, $\eta=12\frac{\mu_l}{\rho_l H^2}$ is the effective drag,
$\mu_l$ the fluid dynamic viscosity, $\rho_l$ its density and $H$ the
height of the channel~\cite{noi_pre,condmat}.  The term $f(z)$ is the
capillary force determined by the wettability properties of the
surface with respect to the fluid, $f(z)= \frac{2 \cos(\theta(z))
  \gamma}{\rho_l H}= 2 \cos(\theta(z))V_{cap} V_{diff} $, being
$\gamma$ the fluid surface tension, $V_{cap}=\gamma/ \mu_l$ the
capillary speed and $V_{diff}=\mu_l/(\rho_l H)$ the diffusive
speed. The angle $\theta(z)$ is usually approximated by the static
contact angle, which depends on the free energy balance of the
solid-liquid-vapor contact line at rest. For the case of moving
fronts, the static contact angle should acquire dynamical corrections
\cite{blake,joanny84}.  However, these corrections are expected to
play a negligible role as compared to contact angle fluctuations
originated by the irregular coating. The two terms on the
left-hand-side of Eq.~\eqref{lw} stem from the time derivative of the
total momentum of the fluid ($d/dt (\rho_l z H \dot{z})$), which
enters the capillary from an infinite reservoir. The dissipative term
on the right-hand-side is due to friction with the walls, which is
proportional to the filled length and to the interface velocity, which
is taken to coincide with the average value of a Poiseuille
transversal velocity profile.

The model of random coating used here consists of a sequence of
patches of length $\Delta$, such that the capillary force is constant
on each patch: $f(z)=f_i$ for $z\in[z_i,z_{i+1}]$, with $z_i=i\Delta$
and $i=0,1,...$. We take $f_i$ to be a random variable with $\langle
f_i f_j \rangle = \langle f_i^2 \rangle \delta_{ij}$.  The probability
density function (pdf) of $f_i$, $P_f(f_i)$, is independent of $i$,
i.e. the random coating is stationary. In a real capillary, the force
$f$ can take values in a bounded interval, being proportional to
$\cos(\theta(z))$: we therefore consider $f_{-}<f_i<f_{+}$, where
$f_{-(+)}= 2 \cos(\theta_{-(+)})V_{cap} V_{diff} $. Since we focus on
filling experiments, such that the initial position of the interface
coincides with the capillary inlet, we also require that $f_{+}>0$.

After a general discussion of
the mathematical properties of Eq.~(\ref{lw}), for illustration purposes,
we shall present explicit calculations for $P_f(f)$ in the case of uniform distribution of the capillary force, although
our analysis is by no means restricted to this specific distribution. 
The present model is inspired to a criterion of maximum simplicity: in particular, the
coating has no long-range correlations (memory is lost above a length
$\Delta$). Despite this simplicity, our model is found to exhibit a 
very rich phenomenology, including power-law tails
in the distribution of waiting times, which closely evokes the stick-slip
behavior observed in recent experiments \cite{schaffer98, schaffer00}.

Upon introducing the dimensionless variables $v=z\dot{z}/(\Delta V_{\Delta})$,
$s=\eta t$ and $g(z)=f(z)/V^2_{\Delta}$, with $V_{\Delta}=\Delta \eta$, Eq.~\eqref{lw} can be 
mapped onto a 'simple' relaxation equation:
\begin{equation} \label{lw2}
\frac{d v}{ds}=-(v-g(z)),
\end{equation}
where, in the following, we shall refer to the variables
$v$ and $g$ as to ``momentum'' and ``force'', respectively. 
The change of variable $f \to g$ defines the boundaries
$g_{-}=f_{-}/V^2_\Delta$ and $g_{+}=f_{+}/V^2_\Delta$, as
well as a transformed pdf $P_g(g)=P_f(f)|df/dg|$. 
Because of the strongly non-linear $z$-dependence of
$g(z)$, Eq.~\eqref{lw2} is very hard to solve 
with the standard analytical tools of the theory of stochastic processes.
However, this equation permits to glean useful information on the
local interface dynamics, i.e. when $z\in[z_i,z_{i+1}]$, so that $g(z)$ is
constant. In particular, the following questions naturally arise: given
the front at position $z_i$, with a given momentum $v$, what is the probability
for the front to advance to $z_{i+1}$? And, what are the corresponding ``waiting time''
$\tau$ and velocity $v'$, once the next location $z_{i+1}$ is reached?  

For any value of $g$, the answers to these questions are exactly determined. 
Given that $g$ is a random variable, however,
$\tau$ and $v'$ also inherit a stochastic character, the corresponding probability distributions
being denoted as $P_\tau(\tau|v,i)$ and $P_{\delta v}(\delta v|v,i)$, where $\delta v=v'-v$ 
is the momentum change upon crossing the patch of length $\Delta$.  
Since $\delta v$ is the increment of $v$ and $\tau$ is the increment of $s$, de-facto,
these conditional distributions represent local propagators for the paths $v(i)$
and $s(i)$, $i$ being the front position.

The momentum increment $\delta v$ in crossing the $i$-th patch, characterized by the
constant force $g$ and initial momentum $v$, is given by $\delta v=v'-v=(g-v)\left(1-e^{-\tau}\right)$,
where $\tau$ is determined by solving the ``exit
equation'', obtained from the conditions $z(0)=z_i$, $z(\tau)=z_{i+1}$:
\begin{equation} \label{exit}
g \tau+(v-g)\left(1-e^{-\tau}\right)-(i+1/2)=0.
\end{equation}
Equation~(\ref{exit}) connects the four variables $i,v,\tau$ and $g$. Solving
this equation with respect to $\tau$, defines a function
$\tau(v,g;i),$ whose smallest positive real solutions represent the
waiting time to go from location $i$ to $(i+1)$ for a fixed $g$ and
$v$. In particular, given the two force-extrema $g_{-}$ and
$g_{+}$, it is possible to define the maximum
($\tau_{max}$) and minimum ($\tau_{min}$) waiting times, to exit 
the $i$-th patch. However, since $\tau(v,g;i)$ is a transcendental function, these
expressions can only be obtained numerically.
\begin{figure}
\begin{center}\includegraphics[width=8cm,clip=true]{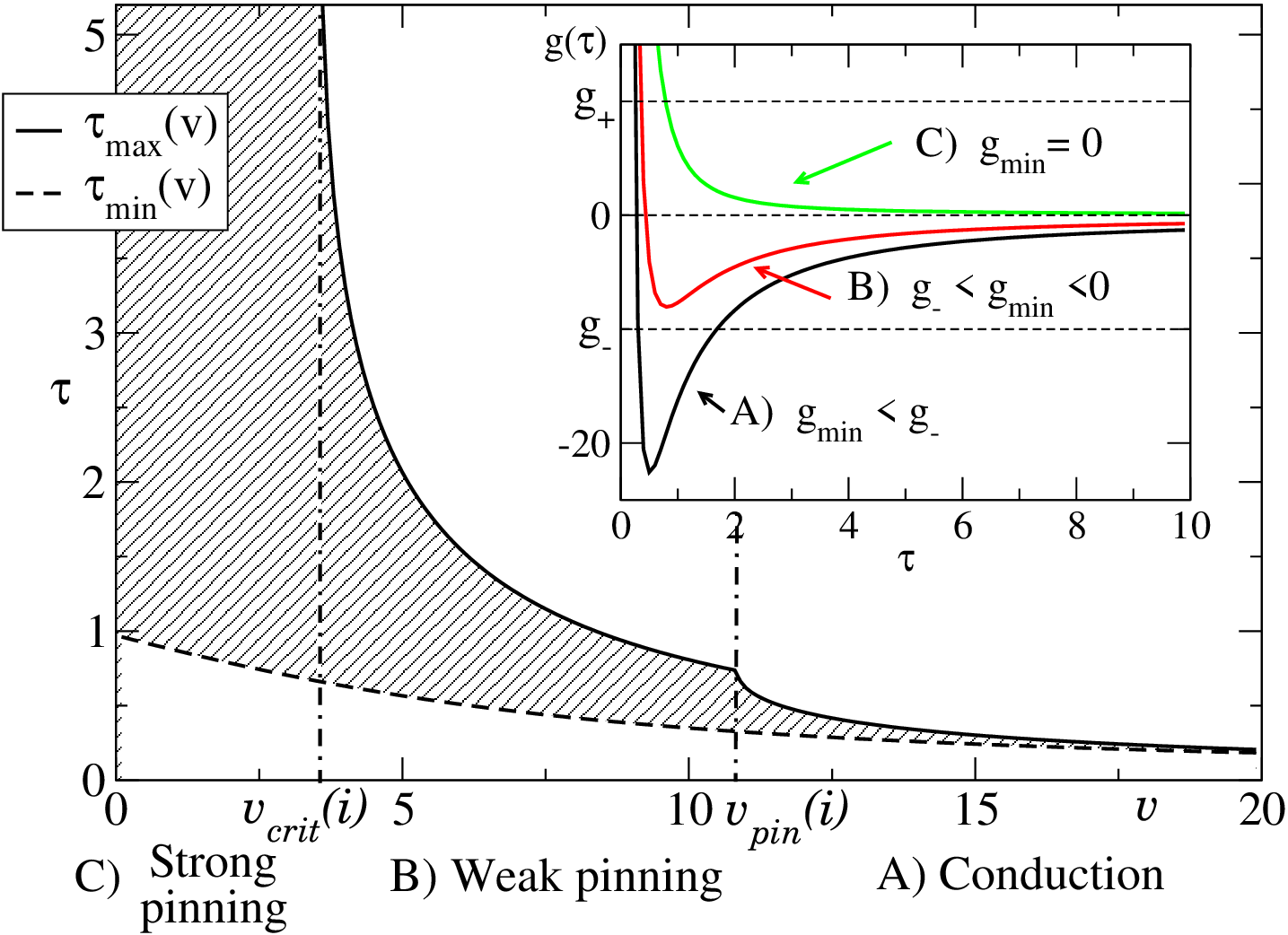}\end{center}
\caption{Minimum and maximum values of the waiting time
  $\tau$ as function of $v$, for $i=3$, with $g_{-}=-10$
  and $g_{+}=10$. Inset: $g(\tau,v;i)$ vs. $\tau$, for $i=3$ and three
  different values of $v$ ($v$=2 for case C, $v=9$ for case B, and $v=13$ for case A).\label{fig:ranges}}
\end{figure}
On the other hand, Eq.~\eqref{exit} is easily inverted with respect to
$g$, determining the function $g(\tau,v;i)=\frac{ (i+1/2- v)+
 v  e^{-\tau }}{\tau -1+e^{-\tau }}$, which is central to our
discussion. In particular, given $i$ and $v$, there exists a threshold
$g_{min}(v;i)=\underset{\tau>0}{min}\;g(\tau,v;i) \le 0$ marking the
minimum value of the force such that the front is guaranteed to reach
position $z_{i+1}$. For values of the force smaller or equal than
$g_{min}(v;i)$, the front is considered ``pinned''. The actual behavior
of $z(t)$ after such a pinning event is a damped
oscillation around a position $z_j$ with $j \le i$ (likely close to
$i$).

At a given choice of $g_{-}, g_{+}$ and $i$, one identifies three
different possible situations, illustrated in Fig.~\ref{fig:ranges},
depending on the initial momentum $v$: A) a "conductive phase",
characterized by $v>v_{pin}(i)$; B) a "weak pinning phase", where $v$
takes intermediate values, $v_{crit}(i)<v<v_{pin}(i)$; C) a "strong
pinning phase'', characterized by low values of momentum, i.e.
$v<v_{crit}(i)$.  Here the value $v_{pin}(i)$ is obtained by inverting
the relation $g_{min}(v_{pin};i)=g_{-}$, and represents the maximum momentum
such that pinning is possible, while $v_{crit}(i)\equiv(i+1/2)$ is the
critical value below which the front is pinned for any non-positive
force.
\begin{table}[htb]
\begin{tabular}{c|c|c|c}
\                   &A                          &B                           &C \\\hline 
$v$                 &$v>v_{pin}(i)$   &$v_{crit}(i)<v<v_{pin}(i)$          &$0<v<v_{crit}(i)$\\\hline 
$g_{min}(v;i)$      &$g_{min}\le g_{-}$              &$g_{-}<g_{min}<0$           &$g_{min}=0$\\\hline 
$\tau_{min}(v;i)$   &$\tau(v,g_{+};i)$        &$\tau(v,g_{+};i)$         &$\tau(v,g_{+};i)$\\\hline 
$\tau_{max}(v;i)$   &$\tau(v,g_{-};i)$        &$\tau(v,g_{min};i)$         &$\infty$\\\hline 
$p_{pin}(v,i)$      &0                          &$\tt{prob}(g\le g_{min})$   &$\tt{prob}(g\le 0)$
\end{tabular}
\caption{Parameter ranges characterizing the three regimes $A$,
$B$ and $C$ and corresponding ranges of the waiting time $\tau$ and pinning
probability $p_{pin}$.  
\label{table}}
\end{table}

Each of these three situations corresponds to a distinct behavior of
the function $g(\tau,v;i)$ (see the inset of Fig.~\ref{fig:ranges}),
which is reflected into different ranges of existence of the waiting
time $\tau$: in the cases A) and B), $\tau$ is bounded both from above
and below, while in the case C) it is bounded only from below.
Figure~\ref{fig:ranges} shows $\tau_{min}(v;i)$ and $\tau_{max}(v;i)$
for two values of $i$ and a choice of $g_{-}$ and $g_{+}$.
The three possible shapes
of $g(\tau,v;i)$ govern also the pinning probability $p_{pin}(v;i)$:
when momentum drops below the value $v_{pin}(i)$, the interface jumps
from regime A to regime B and the pinning probability $p_{pin}(v;i)$
goes from $0$ to a finite value. Further decreasing $v$, the pinning
probability increases.  When momentum $v$ goes below the value
$v_{crit}(i)$, the interface enters the regime C, where the pinning
probability $p_{pin}(v;i)=p^{max}_{pin}=-g_{-}/(g_{+}-g_{-})$, is at
its maximum.  Ranges for $v$, $g_{min}$ and $\tau$, as well as pinning
probabilities, are summarized in Table~\ref{table}.  Translated back
to physical variables, the condition for the phase C, $v \le
v_{crit}(i)$, reads, at large $i$, as $\dot{z} \le V_\Delta$.

\begin{figure}
\begin{center}\includegraphics[width=8cm,clip=true]{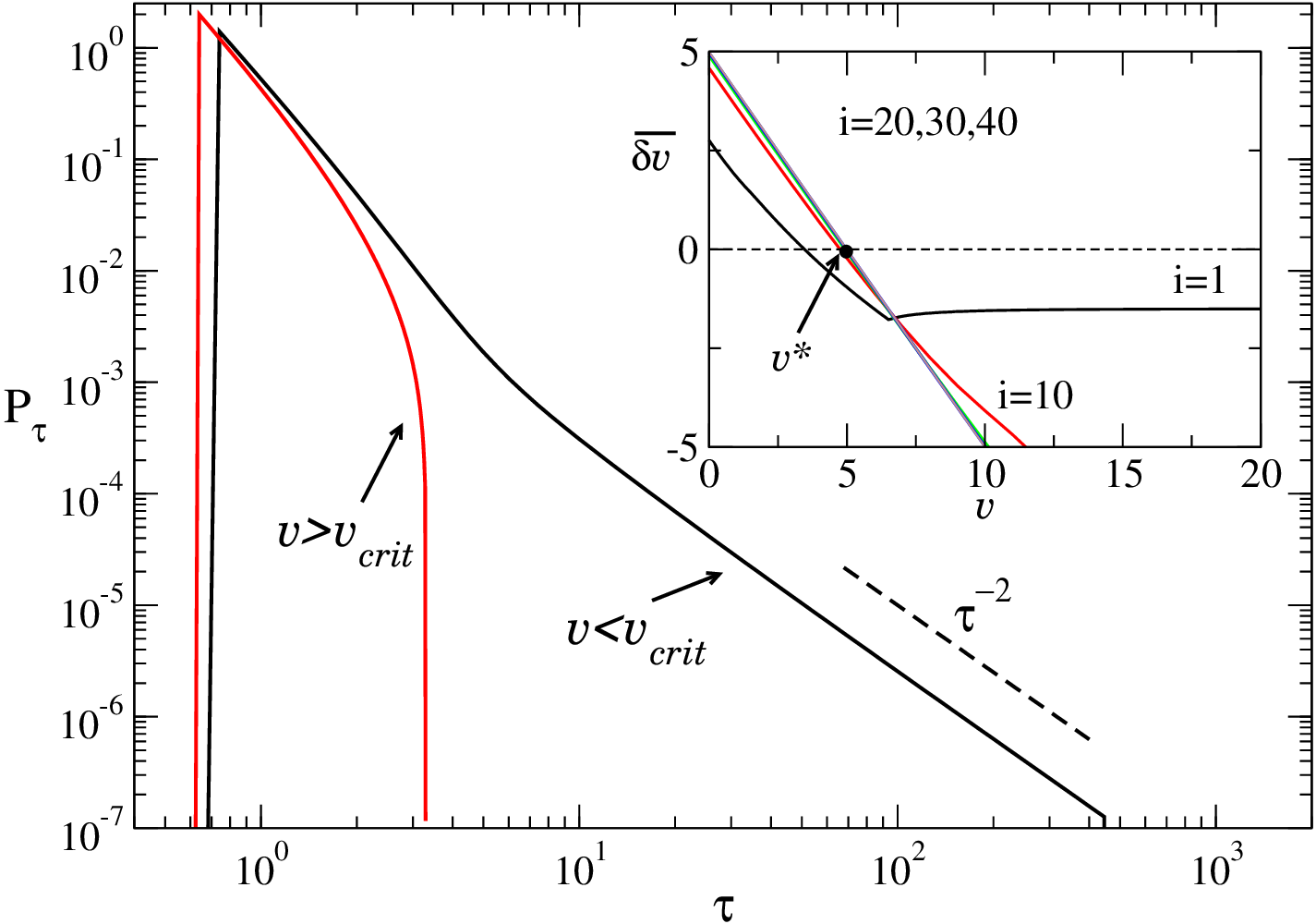}\end{center}
\caption{Probability density function of waiting times $P_\tau$
  vs. $\tau$ for $i=3$ and two values of $v=3<v_{crit}(i)$ and
  $v=4>v_{crit}(i)$, with $g$ uniformly distributed in $[-10,10]$.
  Inset: the average increment $\overline{\delta v}(v,i)$ as a
  function of $v$, for different values of $i$ ($1$, $10$, $20$, $30$
  and $40$). The dramatic emergence of a long-tail for the sub-critical
  case $v<v_{crit}$ is clearly visible. \label{fig:timepdf}}
\end{figure}

The conditional pdf of the waiting times
$P_\tau(\tau|v,i)$ is obtained from the pdf of the force,
$P_g(x)=V^2_{\Delta} P_f(x)$, through the formula $P_\tau(\tau|v,i)=P_g[g(\tau,i,v)]J(\tau,v;i)$,
where
\begin{equation}
J(\tau,v;i)=  \frac{\left|(v_{crit}(i)- v)+ e^{-\tau } [(\tau +1)
   v-v_{crit}(i)]\right|}{\left[(\tau -1)+e^{-\tau}\right]^2}
\end{equation}
is the Jacobian $|\frac{dg(\tau,i,v)}{d\tau}|$. Note that to obtain
the bulk of $P_ \tau$ one does not need the solution $\tau(v,g;i)$ of the transcendental
Eq.~\eqref{exit}. This quantity is however needed to retrieve the
boundaries $\tau_{min}$ and $\tau_{max}$. Note also that, when integrating between $\tau_{min}$ and $\tau_{max}$,
$P_\tau(\tau|v,i)$ is not normalized to $1$, but to $1-p_{pin}(v;i)$.
If the interface is in phase C, the maximum waiting time is
infinite: in this case one sees that $J \sim (v_{crit}(i)-v)\tau^{-2}$
for $\tau \to \infty$. Diverging waiting times correspond to
vanishing values of the force $g \to 0^+$. These two
observations sum up to yield a power-law tail for the waiting time pdf
$P_\tau \sim P_g(0)(v_{crit}(i)-v)\tau^{-2}$: all moments (including
the average) are divergent for this distribution. Such result is even
more remarkable since it does not depend on the precise pdf of the
force $P_f$, provided that $P_f(0^+)>0$, i.e.  arbitrarily small
positive values of $f$ are allowed.  In Fig.~\ref{fig:timepdf}, the
pdfs of $\tau$ for $i=3$ and two values of $v=3<v_{crit}(i)$ and
$v=4>v_{crit}(i)$, are shown. In an experiment with a randomly coated
wall, as soon as the interface velocity $\dot{z}$ drops below
$V_{\Delta}$, we expect to observe strongly fluctuating waiting times,
with possible ``apparent'' pinning events, i.e. situations where the
interface remains stuck for very long times before starting again with
a finite velocity. Assuming $\Delta/H\leq 0.1$, $\nu_l\simeq
10^{-6}m^2/s$ and $H=10^{-6}m$, we obtain $V_\Delta\leq 0.1m/s$, which
appears to be relevant to current experimental conditions
\cite{schaffer98,schaffer00}.

We continue our discussion by considering the pdf of the momentum
increments  $P_{\delta  v}(x|v,i)=\int d\tau
P_\tau(\tau|v,i)\delta[x-\delta v(\tau,v,i)]$.
Numerical inspection
of the analytic properties of the moment-generating function 
$w(\lambda|v,i)=\int  d\tau
P_\tau(\tau|v,i) e^{\lambda \delta v(\tau,v,i)}$ reveals that
$P_{\delta v}(\delta v|v,i)$ always has finite moments, even when $v$
drops below $v_{crit}(i)$.
The conditional average increment, restricted to the ensemble of unpinned 
trajectories, reads
\begin{equation}
\label{dv}
\overline{\delta v}(v,i)=\int_{\tau_{min}(v,i)}^{\tau_{max}(v,i)} d\tau P^*_\tau(\tau|v,i) \delta v(\tau,v,i),
\end{equation}
where
$P^*_\tau(\tau|v,i)=P_\tau(\tau|v,i)/(1-p_{pin}(v,i))$. This quantity is shown in
the inset of Fig.~\ref{fig:timepdf} for  different values of $i$. 
It is interesting to note that, for large
values of $i$, $\overline{\delta v}$ no longer depends on $i$ and goes 
linearly with $v$, $\overline{\delta v} \sim (v^*-v)$, with $v^*$ a constant. 
This can be explained by assuming that, at large $i$, the interface is on average in the
strong-pinning phase C, $\tau \sim \infty$, so that $\delta v \approx
g-v$, i.e. $v^*=\int_0^{g_{+}} g P_g(g)$. 
The  surviving (unpinned) trajectories, tend to cluster on a constant momentum 
ensemble, $z_i\dot{z}_i=v^*
\Delta^2 \eta$. 
\begin{figure}[htb]
\begin{center}
\includegraphics[width=9cm,clip=true] {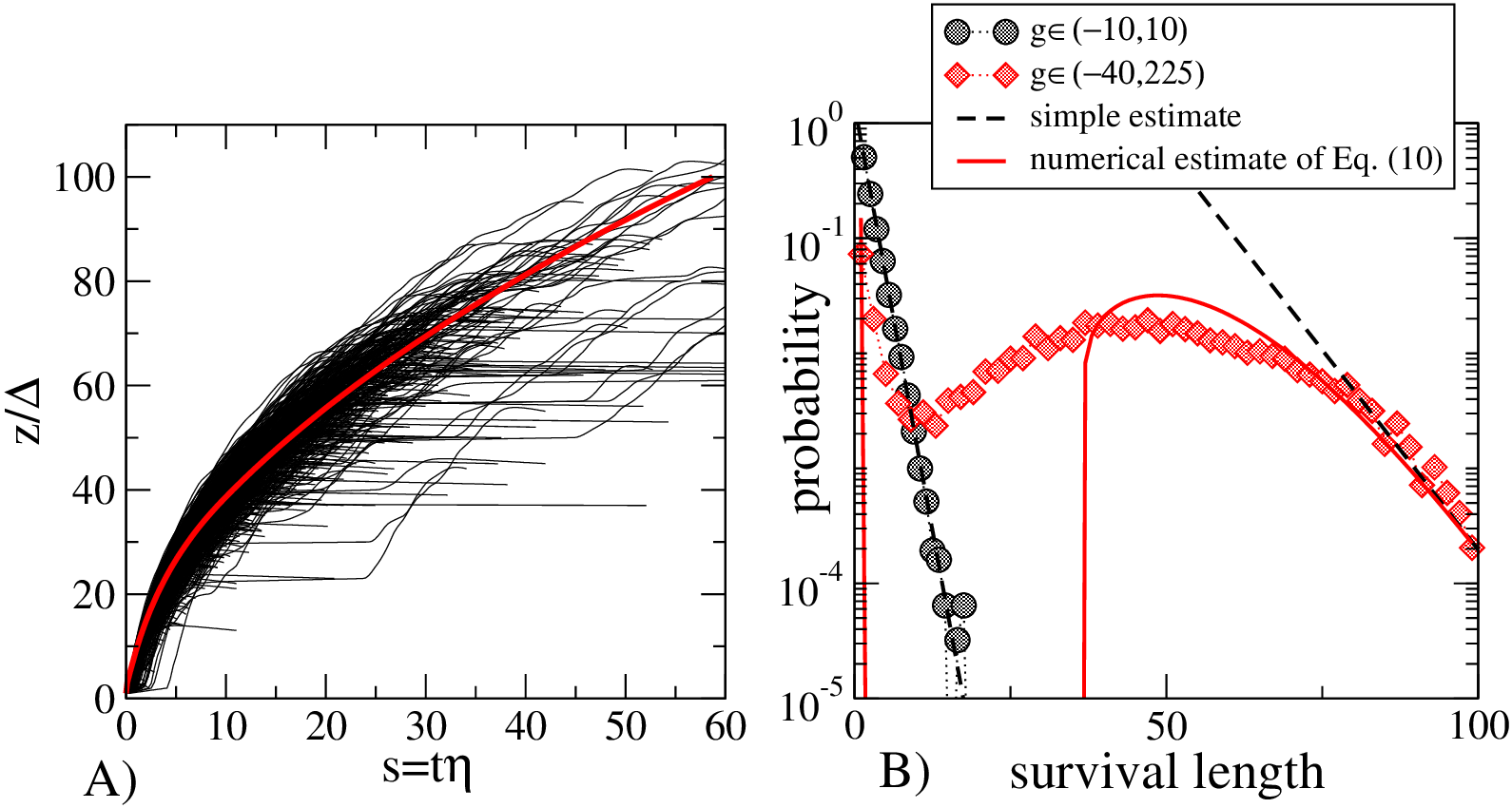}
\end{center}
\caption{Left) An ensemble of $1000$ trajectories as obtained from the
numerical simulations of Eq.~(\ref{lw}), and the average trajectory
$z(t)/\Delta$ vs.  $\eta t$, (solid line in the middle), obtained as
described in the text (see Eq.~(\ref{dv})).  Right) Distribution of
the pinning lengths; comparison between simulations (data points), the
numerical estimate of Eq. (6) (solid line) and the theoretical
estimate $e^{-qi}$, as given in the text (dashed line).\label{fig:pin}}
\end{figure}

Equation~\eqref{lw} can be numerically integrated in order to gather a
large statistics with many realizations of the random coating. This
may take large computational time, mostly due to trajectories
entering regime $C$, which can spend a long time in the
same coated patch. A better alternative is to resort to numerical calculations
based upon our analytical expressions, which can yield
average quantities of interest for experimental or industrial design.
Starting at $i=1$ with a given initial value of the fluid momentum,
e.g. $v=0$, one can iteratively generate a ``mean'' trajectory
$\overline{v}(i)=\sum_{j=1}^i\overline{\delta
  v}(\overline{v}(j-1),j)$. We observe that this yields a good
estimate of the average of the surviving trajectories $\langle v(i)
\rangle$, which, in the original variables, corresponds to $\langle
\dot{z} \rangle$ as a function of $z$. Such observation ensures that
$P_{\delta v}$ remains reasonably peaked, i.e. that the
mean increment provides a fair estimate of the interface motion. 

We can next compute the average waiting time $\overline{\tau}(i)=\int dx
P_\tau(x|\overline{v}(i),i) x$, based upon the aforementioned mean value of the
momentum $\overline{v}(i)$. Again, by summing up all
average waiting times, one recovers an estimate for the total elapsed
time $\overline{s}(i)=\sum_{j=1}^i \overline{\tau}(j)$. 
A satisfactory agreement between $i(\overline{s})$ and a large sets of trajectories
obtained by numerical integration of Eq.~\eqref{lw}, is shown in
Fig.~\ref{fig:pin}A. Note that this procedure makes sense only as long as 
$\overline{\tau}(i)$ is well defined, i.e. until
$\overline{v}(i)>v_{crit}(i)$. As one can see, in Fig.~\ref{fig:pin}A
some of the simulated trajectories drop into this low-momentum state,
and very long waiting times are observed indeed, in the form
of long plateaux of $z(t)$, before the interface starts moving again.
This corresponds to a sort of stick-slip behavior for the front
dynamics, very similar to the one reported in Fig. 2c of
Ref.\cite{schaffer98}.  
Typical orders of magnitude under experimental conditions are:
$V_{cap}\simeq 100m/s$, $V_{diff}\simeq 10^{-1}m/s$,$V_{\Delta}\leq
0.1m/s$, yielding $g \simeq 10^3 \cos(\theta)$. 
In Fig.~\ref{fig:pin}A we have taken $\delta g=g_+-g_-\simeq 300$, corresponding
to fluctuations of the contact angle $\delta \cos(\theta)
\simeq 0.3$.

From the average trajectory $\overline{v}(i)$, an ``average''
conditional pinning probability $p_{pin}(\overline{v},i)$ at each
position $i$ can also be computed. The total probability of observing 
front pinning at position $i$ is given by the product of the probability of
not being pinned at all locations $j<i$ and the one of being pinned at $j=i$, that is:
\begin{equation} \label{numerical}
P_{pin}(i)=\left ( \prod_{j=1}^{i-1}(1-p_{pin}(\overline{v}(j),j)) \right )\;p_{pin}(\overline{v}(i),i).
\end{equation}
This quantity is numerically computed and compared with the pdf of the
pinning lengths, as obtained from direct numerical simulations of
Eq.~\eqref{lw} (see Fig.~\ref{fig:pin}B). The theoretical curve
underestimates the pinning probability during the conductive phase,
however, the rest of the pdf is well reproduced, including the initial
peak due to the $v=0$ starting condition.

A simpler prediction, which does not require any numerical computation
of Eq.~\eqref{numerical}, can be obtained from the observation that
$\overline{v(i)} \to v^*$ for $i >> 1$. When this saturation value
meets the critical line $v_{crit}(i) \approx i$, i.e. when $i \sim
v^*$, the pinning probability at each new patch is simply given by the
maximum probability $p^{max}_{pin}$.  From there on, pinning is just
one of the two possible outcomes of a Bernoulli process, implying that
the survival probability decays exponentially, $P_{pin}(i) \sim
\exp(-q i)$ with $q=\log(1-p^{max}_{pin})$. This exponential tail,
which marks the ``end'' of the capillary filling, for a uniform
$P_f(f)$, begins at position $i\simeq g_{+}/2 = \cos(\theta_+) V_{cap}
V_{diff}/V^2_{\Delta}$.  These simple expressions may offer a handy
way to estimate the pinning length under experimental conditions.

Summarizing, we have highlighted the
non-trivial properties of the Lucas-Washburn
equation~\eqref{lw} with a random capillary force.
Our analysis unveils the presence of a regime characterized by
a broad distribution of waiting times $P_\tau \sim \tau^{-2}$ (if $f_- \le 0$). 
The same analysis also permits to develop qualitative estimates of the
maximal pinning length $z_{max}/\Delta \sim \cos(\theta_+) V_{cap} V_{diff}/V^2_{\Delta} $ 
as well as the slope of its exponential distribution tail $q \sim \log
[f_+/(f_+-f_-)]$ (if $f_- <0$). 
Finally, we have developed a fast numerical procedure to retrieve detailed information,
such as a more complete estimate of the pinning length distribution or
of the average space-time trajectory inside the channel, given the
statistical properties of the surface. 
These results could also be exploited in reverse, i.e. inferring information about the
wall roughness statistics by performing many filling experiments on different capillaries.

\section{Acknowledgments}
We wish to thank S. Chibbaro for useful discussions.  Financial support
through the NMP-031980 EU project (INFLUS) is kindly acknowledged.

\end{document}